\documentclass[proof]{pasj00}
\usepackage{color}

\begin{document}
\SetRunningHead{Hirota et al.}{H$_{2}$O Supermaser Burst in Orion~KL}

\title{VERA and ALMA Observations of the H$_{2}$O Supermaser Burst in Orion~KL}

\author{
Tomoya \textsc{hirota}\altaffilmark{1,2}
Masato \textsc{tsuboi}\altaffilmark{3}
Yasutaka \textsc{kurono}\altaffilmark{4,5}
Kenta \textsc{fujisawa}\altaffilmark{6,7} 
Mareki \textsc{honma}\altaffilmark{1,2} 
Mi Kyoung \textsc{kim}\altaffilmark{8}
Hiroshi \textsc{imai}\altaffilmark{9}
and 
Yoshinori \textsc{yonekura}\altaffilmark{10}
}
\altaffiltext{1}{National Astronomical Observatory of Japan, Mitaka, Tokyo 181-8588}
\altaffiltext{2}{Department of Astronomical Sciences, The Graduate University for Advanced Studies (SOKENDAI), Mitaka, Tokyo 181-8588}
\email{tomoya.hirota@nao.ac.jp}
\altaffiltext{3}{Institute of Space and Astronautical Science, Japan Aerospace Exploration Agency, Sagamihara, Kanagawa 229-8510}
\altaffiltext{4}{Chile Observatory, National Astronomical Observatory of Japan, Osawa 2-21-1, Mitaka, Tokyo 181-8588}
\altaffiltext{5}{Joint ALMA Observatory, Alonso de Cordova 3107 Vitacura, Santiago 763-0355, Chile}
\altaffiltext{6}{Department of Physics, Faculty of Science, Yamaguchi University, Yamaguchi, Yamaguchi 753-8512}
\altaffiltext{7}{The Research Institute of Time Studies, Yamaguchi University, Yamaguchi, Yamaguchi 753-8511}
\altaffiltext{8}{Korea Astronomy and Space Science Institute, Hwaam-dong 61-1, Yuseong-gu, Daejeon, 305-348, Republic of Korea}
\altaffiltext{9}{Graduate School of Science and Engineering, Kagoshima University, Kagoshima, Kagoshima 890-0065}
\altaffiltext{10}{Center for Astronomy, Ibaraki University, Mito, Ibaraki 310-8512}


\maketitle

\begin{abstract}
The 22~GHz H$_{2}$O maser in Orion~KL has shown extraordinary burst events in 1979-1985 and 1998-1999, sometimes called supermaser. 
We have conducted monitoring observations of the supermaser in Orion~KL using VERA, VLBI Exploration of Radio Astrometry, in the current third burst since 2011 March. 
Three flux maxima are detected in 2011 and 2012 with rising and falling timescales of 2-7~months. 
Time variations of the supermaser seem symmetric for all of the active phases. 
The maximum total flux density of 135000~Jy is observed in 2012 June while it is still one order of magnitude lower than those in previous bursts. 
The supermaser consists of two spatially different components at different velocities. 
They are elongated along a northwest-southeast direction perpendicular to the low-velocity outflow driven by Source~I. 
Proper motions of the supermaser features with respect to Source~I are measured toward west and southwest directions, almost parallel to the low-velocity outflow. 
The flux density and linewidth show an anti-correlation as expected for an unsaturated maser emission. 
The supermaser is located close to the methylformate (HCOOCH$_{3}$) line and continuum emission peaks in the Orion Compact Ridge detected by ALMA, Atacama Large Millimeter/Submillimeter Array. 
The broader velocity range of the weak HCOOCH$_{3}$ emission at the supermaser position would be an evidence of a shock front. 
On the other hand, the 321~GHz H$_{2}$O line is not detected at the position of the supermaser. 
It can be explained qualitatively by one of the theoretical H$_{2}$O excitation models without extraordinary conditions. 
Our results support a scenario that the supermaser is excited in the dense gas interacting with the low-velocity outflow in the Compact Ridge. 
The extremely high flux density and its symmetric time variation for rising and falling phases could be explained by a beaming effect during the amplification process rather than changes in physical conditions. 
\end{abstract}

\section{Introduction}

Orion~KL (Kleinmann-Low object) is one of the nearest massive star-forming regions at a distance of 420~pc \citep{menten2007, kim2008} and hence, has been the most extensively studied source in the context of star-formation study \citep{genzel1989, bally2008}. 
There are powerful outflows and jets possibly associated with young stellar objects (YSOs) in this region at a wide range of spatial scales from $<$100~AU to $>$10000~AU. 
They are traced by various molecular lines such as vibrationally excited SiO masers at 43~GHz \citep{kim2008, matthews2010,kim2014}, 22~GHz H$_{2}$O maser \citep{genzel1981,gaume1998,hirota2007,greenhill2013,kim2014}, millimeter/submillimeter thermal SiO \citep{plambeck2009, zapata2012, niederhofer2012} and CO lines \citep{beuther2008, zapata2009}, and near-infrared H$_{2}$ line \citep{allen1993, kaifu2000, bally2011}. 
These observational studies have revealed physical properties of the outflows, jets, and surrounding shocked materials interacting with ambient clouds. 

Among a number of tracers, the 22~GHz H$_{2}$O maser is known to be a unique tracer to investigate a three-dimensional velocity structure. 
Thanks to its extremely high brightness, it can be detected by high-resolution interferometers and very long baseline interferometers (VLBI) providing a capability of proper motion measurements. 
As a result, the 22~GHz H$_{2}$O maser in Orion~KL is found to be tracing a low-velocity outflow along a northeast-southwest axis \citep{genzel1981,hirota2007,greenhill2013,kim2014} centered at a strong radio continuum source labeled I \citep{churchwell1987, menten1995}. 
Hereafter we call this Source~I. 

One of these H$_{2}$O maser features in Orion~KL has exhibited an enormous outburst, which is sometimes called supermaser. 
The H$_{2}$O maser burst in Orion~KL was reported in 1979-1985 for the first time \citep{matveenko1988, garay1989} followed by 1998-1999 \citep{omodaka1999, matveyenko2004, shimoikura2005}. 
Based on VLBI observations, the supermaser was revealed to have east-west or northwest-southeast elongations in 1979-1985 and 1998-1999, respectively. 
The line-of-sight velocity with respect to the local standard of rest (LSR) was stable at $\sim$ 8~km~s$^{-1}$, suggesting a common origin. 
Although the maser burst could be related to star-formation activities in Orion~KL, 
either a circumstellar disk \citep{matveenko1988,shimoikura2005} or a jet/outflow driven by a YSO \citep{garay1989,matveyenko2004}, its origin and mechanism are still under debate. 

The third burst of the H$_{2}$O maser in Orion~KL has started since February 2011 \citep{tolmachev2011, otto2012, matveyenko2012a, matveyenko2012b}. 
The burst appears to be possibly periodic with an interval of about 13~years (1985, 1998, and 2011). 
It could support the hypothesis of a common origin \citep{tolmachev2011}. 
We have carried out astrometric monitoring observations with VLBI Exploration of Radio Astrometry (VERA) since 2011 March \citep{hirota2011} and studied basic properties of the supermaser. 
Main conclusions by \citet{hirota2011} are (1) the burst occurs at two spatially nearby but distinct components (separation of $\sim$10~mas) at the LSR velocities of 6.95 and 7.58~km~s$^{-1}$ (hereafter we call them "6.9~km~s$^{-1}$" and "7.5~km~s$^{-1}$" features, respectively, using representative velocities) as reported in previous bursts \citep{matveenko1988, garay1989, omodaka1999, matveyenko2004, shimoikura2005}, (2) these features are elongated along a northwest-southeast direction perpendicular to the low-velocity outflow and their proper motions are parallel to the outflow axis, and (3) the absolute position of the supermaser is coincident with a shocked molecular gas called the Orion Compact Ridge. 
All of these characteristics suggest that the supermaser could be excited in an interacting region between the low-velocity outflow and the ambient gas in the Compact Ridge \citep{genzel1981, liu2002, favre2011, brouillet2013}. 

Nevertheless, the maximum flux density for the current burst phase is reported to be 44000~Jy on May 2011 \citep{hirota2011}. 
It is much lower than those in previous bursts in which the flux densities of the supermaser reached 10$^{6}$~Jy \citep{matveenko1988, garay1989, omodaka1999, matveyenko2004, shimoikura2005}. 
Although the maser bursts could have possible periodicity of 13~years, 1985, 1998, and 2011, it is still unclear whether all of three events are caused by a common origin. 
In order to understand their origin and to reveal their relationship between star-formation activities in Orion~KL in relation to outflows, jets, and interaction with ambient clouds, 
it is important to continue further monitoring observations of the supermaser at a current valuable opportunity. 

In this paper, we report results of our continued monitoring observations of the supermaser with VERA to investigate time variations of their properties. 
In addition, we present follow-up observations with Atacama Large Millimeter/Submillimeter Array (ALMA) to identify possible powering source(s), as well as to study pumping mechanisms based on multi-transition H$_{2}$O lines. 
Although we find that the current burst activity is lower by a factor of 10 than those in previous bursts in terms of the maser flux density as discussed later, we call the current bursting maser feature "supermaser" throughout the present paper. 

\section{Observations}

\subsection{VERA}

Observations of the $6_{1,6}$-$5_{2,3}$ transition of H$_{2}$O at 22.235080~GHz \citep{kukolich1969, chen2000} in Orion~KL were carried out with VERA. 
Details of the observations have been described in separate papers \citep{hirota2007, hirota2011}, and we briefly summarize key parameters in this section. 
VLBI monitoring observations have been conducted regularly since 2011 March with a typical interval of 1-4~weeks. 
The maximum baseline length of VERA is 2270~km and the resultant fringe-spacing is 1.2~mas. 
We employed the dual beam observation mode, in which Orion~KL and an ICRF source J054138.0-054149 \citep{ma1998} were observed simultaneously, to measure the absolute position of the target H$_{2}$O maser in Orion~KL with respect to the positional reference source. 
The astrometric accuracy is estimated to be 0.4~mas according to post-fit residuals of proper motion measurements. 
These two sources were observed for 8~hours being tracked from horizon to horizon at the lowest elevation angle of $\sim$15~degrees. 
A typical on-source time was 6~hours for each epoch. 
In this paper, we report the astrometric and imaging results of the first 18 epochs of observations from 2011/068 to 2012/269 (hereafter we indicate an epoch of observation as year/day of the year). 
A spectral resolution was set to be 15.625~kHz, corresponding to a velocity resolution of 0.21~km~s$^{-1}$. 
Calibration and synthesis imaging were performed using the NRAO Astronomical Image Processing System (AIPS) software package in the same manner employed previously \citep{hirota2007, hirota2011}. 

We also used other data of VERA observations at 22~GHz including Orion~KL as a fringe-finder to investigate a time variation in spectral profiles. 
A total number of spectral line monitoring data is 53 epochs from 2008/329 to 2014/025. 
On-source integration times of each spectral line monitoring data for Orion~KL was 5-10~minutes. 

\subsection{ALMA}

In addition, we used ALMA data taken in one of the early science projects in cycle~0 (ADS/JAO.ALMA\#2011.0.00199.S) to compare images of submillimeter continuum emissions and molecular lines. 
Observations were carried out at the ALMA band~7 (in the range of 320-340~GHz) on 2012 July 16, August 25, and October 21. 
They were almost coincident with the most active burst phase as discussed in the next section. 
In this paper, we mainly focus on the $10_{2,9}$-$9_{3,6}$ H$_{2}$O line at 321.225656~GHz \citep{chen2000} and methyl formate (HCOOCH$_{3}$) lines close to the H$_{2}$O line. 
The maximum baseline length in our ALMA observations was 400~m, corresponding to a spatial resolution of $\sim$0.5\arcsec, depending on the UV weighting in the imaging. 
Synthesis imaging and self-calibration were done with the Common Astronomy Software Applications (CASA) package. 
Further details of observations are reported elsewhere \citep{hirota2014a, hirota2014b}. 

\section{Results}
\subsection{Time variation of the 22~GHz H$_{2}$O maser}
\label{sec-vera}

\begin{figure*}
\begin{center}
\includegraphics[width=18cm]{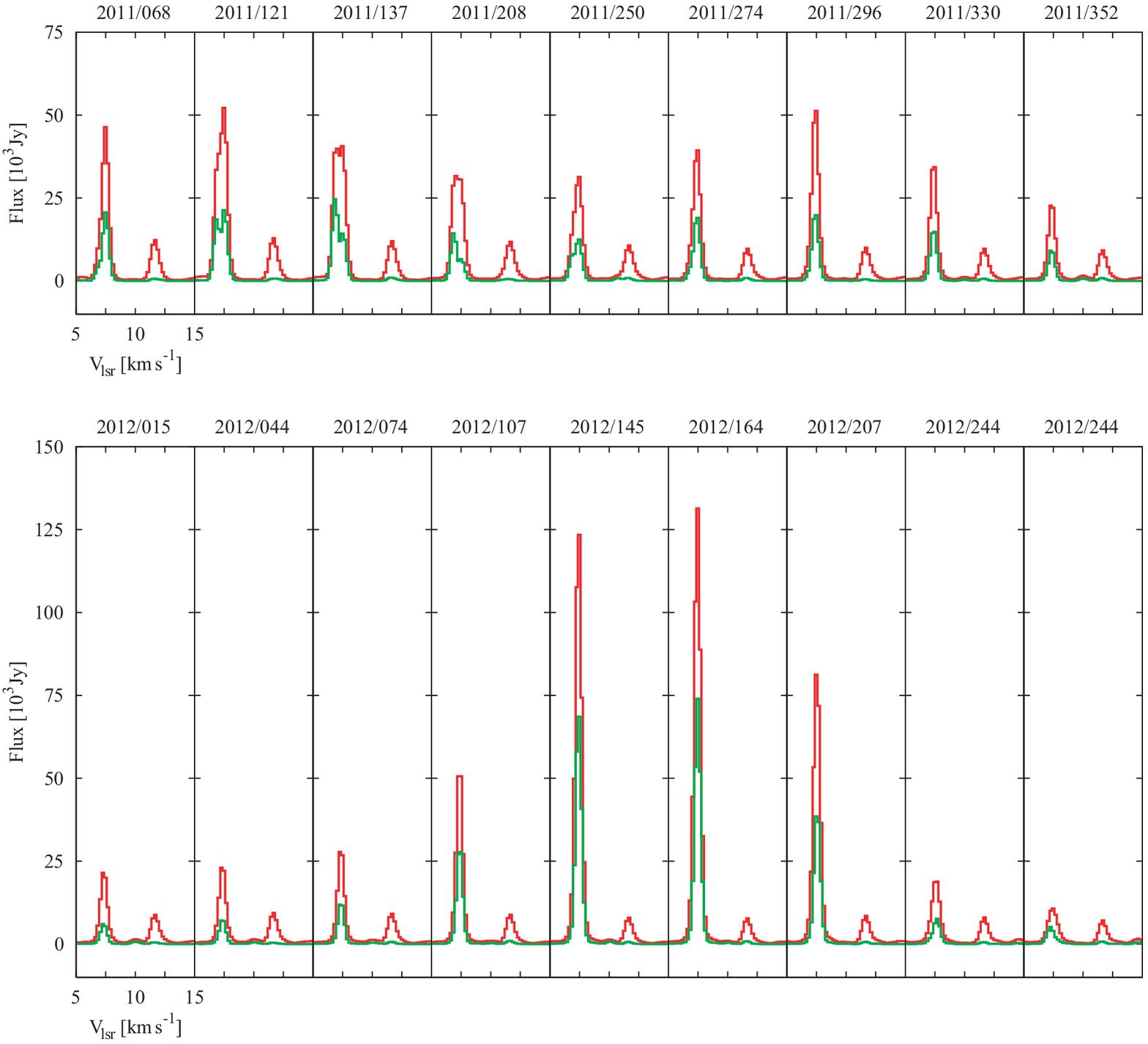}
\caption{Spectra of the supermaser in Orion~KL observed in regular astrometry observations in 2011 and 2012. 
Red and green lines represent the total-power and cross-power spectra, respectively, integrated over all baselines/antennas and all scans. 
The observed epoch is indicated at the top of each panel. 
}
\label{fig-sp}
\end{center}
\end{figure*}

\begin{figure*}
\begin{center}
\includegraphics[width=9cm]{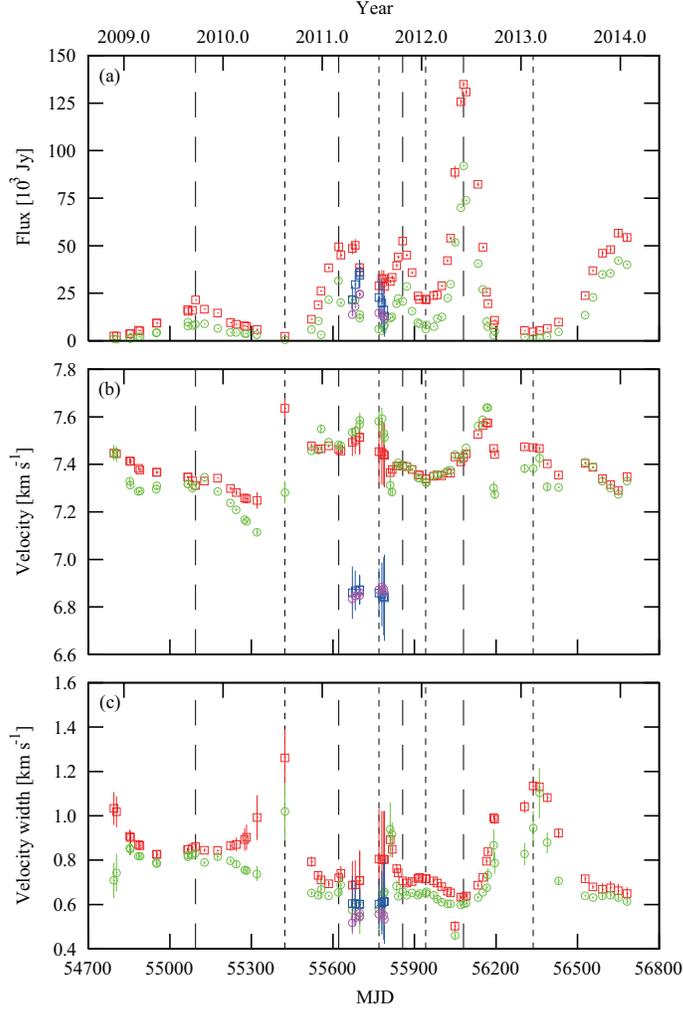}
\caption{Time variations of 
(a) total and correlated flux densities, (b) peak velocity, and (c) velocity width of the supermaser. 
Results of both 18 epochs of regular astrometry and 53 epochs of spectral line monitoring observations are plotted. 
In each panel, results of the Gaussian fitting of the total-power and cross-power spectra are indicated with the $3\sigma$ errors, although the error bars are hardly seen due to high signal-to-noise ratios. 
Squares and circles represent the best-fit parameters for total-power spectra and cross-power spectra, respectively. 
In the middle of 2011, spectra clearly show double-peaked profiles, and hence, two different Gaussian components are fitted to the observed spectra, as shown in red/green and blue/magenta for the $\sim$6.9~km~s$^{-1}$ and 7.5~km~s$^{-1}$ components, respectively. 
Dashed and dotted lines represent epochs with the maximum and minimum flux densities, respectively. 
}
\label{fig-time}
\end{center}
\end{figure*}

Figure \ref{fig-sp} shows spectra of the supermaser observed in 18 epochs of the regular astrometry observations from 2011 to 2012, and Figure \ref{fig-time} shows time variations of the flux density, velocity width, and peak velocity for all the observed epochs including both 18 epochs of the regular astrometry and 53 epochs of the spectral line monitoring observations. 
First, these parameters are determined by fitting a single Gaussian to the observed profile. 
However, there are two distinct velocity components at $\sim$6.9~km~s$^{-1}$ and $\sim$7.5~km~s$^{-1}$ as reported in \citet{hirota2011}. 
These components can be clearly seen in spectral profiles such as the double-peaked structure on 2011/121 and 2011/137 and red-shifted shoulder on 2011/208 (Figure \ref{fig-sp}). 
This results in a significant deviation from the observed line profile in the fitting. 
Thus, we next fit two Gaussians to the observed profiles for these epochs. 
Although these line parameters show slightly larger error bars as shown in Figure \ref{fig-time}, we can see time variations of flux density, velocity width, and peak velocity for the $\sim$7.5~km~s$^{-1}$ component continuously changing even during this phase. 
In contrast, the velocity component that appeared later at $\sim$6.9~km~s$^{-1}$ shows a different time variation. 

According to characteristics of the time variation in Figure \ref{fig-time}, we can divide the monitoring period of the current H$_{2}$O maser burst into four phases; 
a moderately active phase in 2009 followed by a relatively quiescent phase in 2010, a start of the burst with rapid variation in 2011, the maximum phase in 2012, and a quiescent/possible initial phase of the next burst in 2013. 

\begin{figure*}
\begin{center}
\includegraphics[width=18cm]{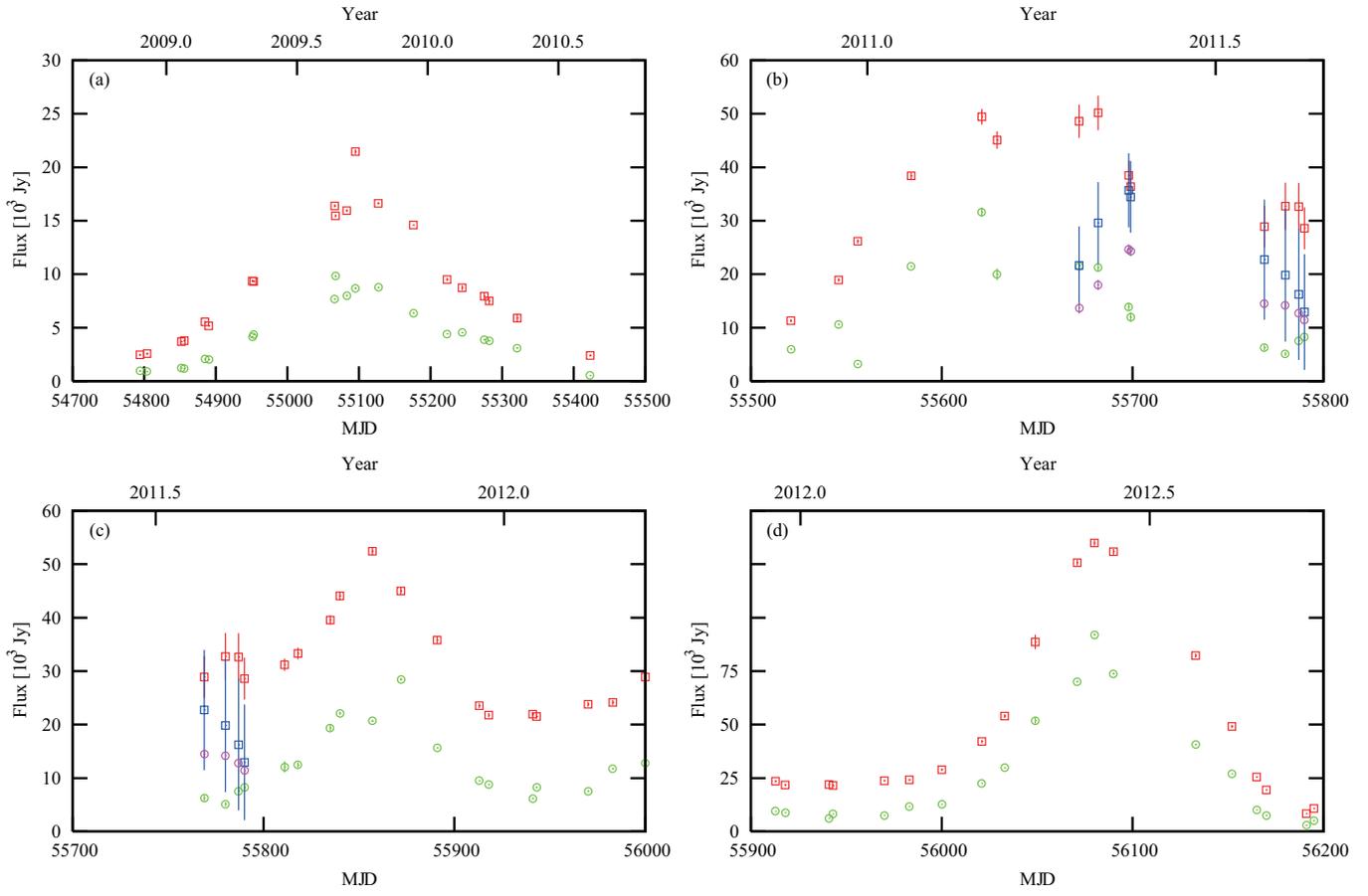}
\caption{Enlarged views of the flux variations for each burst as shown in Figure \ref{fig-time}. 
(a) 2009-2010, (b) first burst in 2011, (c) second burst in 2011, and (d) 2012. 
Symbols are the same as Figure \ref{fig-time}. 
}
\label{fig-flux}
\end{center}
\end{figure*}

In 2009, there is a moderately active phase slowly increasing and decreasing in the flux density both in total power and cross power spectra. 
Timescales of both rising and falling phases, which are defined as periods from minimum to maximum and maximum to minimum flux densities, respectively, are almost equal, 10~months and 11~months, respectively (Figure \ref{fig-flux}(a)). 
During this phase, the peak velocity is also slightly changing from 7.4 to 7.3~km~s$^{-1}$ while the velocity width, $\sim$0.8~km~s$^{-1}$, is almost stable with time. 
After this moderately active phase in 2009, the maser activity becomes a quiescent phase, whereas the total and correlated flux densities are as high as 2400~Jy and 500~Jy, respectively, at the minimum phase on 2010/227. 
Toward this minimum phase, the velocity width of the total-power spectrum gradually increases although that of the cross-power spectrum shows a decreasing trend. 
At the minimum phase on 2010/227, the velocity width and peak velocity change drastically.  Furthermore, these parameters show significant differences between total-power and cross-power spectra but with larger uncertainties than other epochs. 
This could be caused by other velocity components as their contributions become more and more  significant relative to that of the 7.5~km~s$^{-1}$ component. 
Before and after this minimum phase, there were breaks of monitoring observations each for 100~days due to the maintenance of the VERA antennas, and hence, we could not trace time variations in this quiescent phase. 

The H$_{2}$O maser burst has started in the end of 2010 or the beginning of 2011 \citep{tolmachev2011}. 
The flux density of the total-power spectrum at $\sim$7.5~km~s$^{-1}$ has started to increase since 2010/350 while that of the cross power spectrum is temporarily halted on 2010/360 followed by the brightening from 2011/023. 
Although there is a lack of maser images, this might reflect a variation in the spatial structure of the emitting region. 
At the beginning of the current burst phase in 2011, there are two maximum phases on 2011/60-2011/121 and 2011/296-2011/311 with the total flux density of about 50000~Jy (Figure \ref{fig-time}). 
The timescales of rising and falling phases are 7~months and 3~months, respectively, for the maximum phase of 2011/60-2011/121 (Figure \ref{fig-flux}(b)), although the flux density in this phase shows a variability rather than a monotonic increase or decrease. 
The timescales for the subsequent maximum phase of 2011/296-2011/311 (Figure \ref{fig-flux}(c)) are 2~months for both rising and falling phases. 
Apart from the different timescales, the behaviors in rising and falling phases seem to be symmetric as seen in the active phase in 2009 (Figure \ref{fig-flux}). 
As discussed above, the burst appears to occur in two different velocity components at 6.9~km~s$^{-1}$ and 7.5~km~s$^{-1}$ \citep{hirota2011} as seen in Figures \ref{fig-time}(b) and (c). 

The burst activity is temporarily stopped at the end of 2011, but more active burst has started in early 2012. 
The maximum total flux density of 135000~Jy is recorded on 2012/154 at the peak velocity of 7.4~km~s$^{-1}$. 
In this most active phase, the flux density monotonically increases with the timescale of 5~months. 
The velocity width is about 0.60-0.75~km~s$^{-1}$, which is the narrowest value observed in our monitoring as shown in Figure \ref{fig-time}(b). 
On 2012/123, the derived velocity widths show the minimum values of 0.50$\pm$0.02~km~s$^{-1}$ and 0.46$\pm$0.02~km~s$^{-1}$ for the total-power and cross-power spectra, respectively. 
Only in this epoch, the velocity widths are significantly narrower than those observed just before and after this epoch (2012/107 and 2012/145). 
Due to the lack of maser images on 2012/123, we do not discuss what is the true reason for these narrower velocity widths only at this epoch. 
The flux density of the maser decreases with the timescale of 4~months after this maximum phase. 
It should be noted that the time variations in rising and falling phases again seem to be symmetric as seen in Figure \ref{fig-flux}(d). 
At the end of the burst phase in 2012/265 and 2012/269, the peak velocities of cross-power spectra show significant differences from those of the total-power spectra. 
They are thought to be the results of significant contributions from other velocity components and/or effects of structure changes in maser features. 
This effect is seen particularly in less active phases such as in 2010. 

Finally, the burst activity seems to end around 2012/269 while it again appears to be brightening from 2013/138. 
It is most likely an ignition of a newly occurred burst event. 
We will continue monitoring observations with VERA of the time variation of the supermaser in Orion~KL. 
Still, the activity of the 22~GHz H$_{2}$O maser burst in 2011-2012 is relatively lower by a factor of 10 in comparison with the previous bursts in which the peak flux densities reached 10$^{6}$~Jy. 

\begin{figure*}
\begin{center}
\includegraphics[width=9cm]{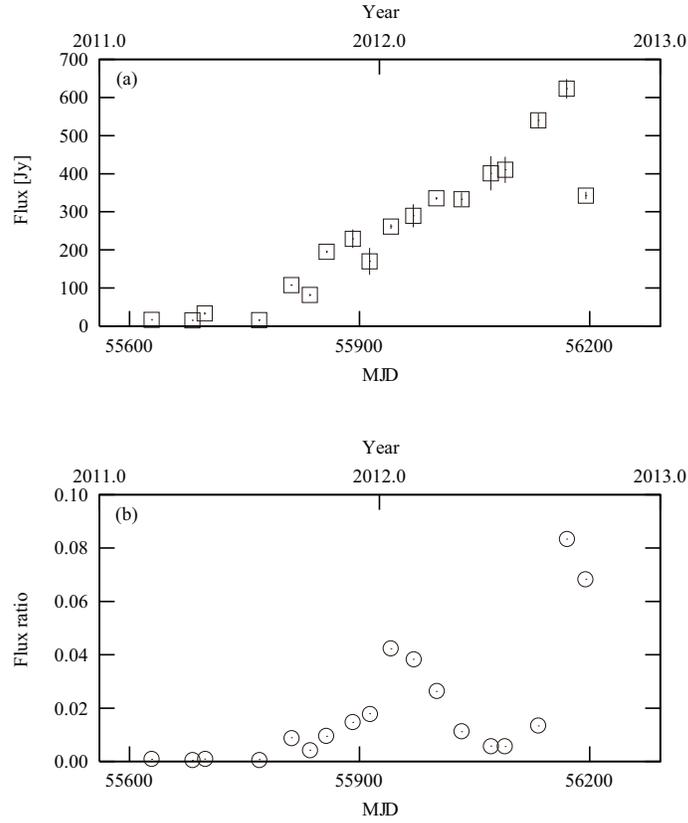}
\caption{Time variation of (a) the flux density of SE feature and (b) the flux density ratio relative to the supermaser feature observed in 2011 and 2012. 
Correlated flux density in Figure \ref{fig-time}(a) is employed to calculate the flux density ratio. 
When there are two velocity components in the supermaser, sum of these flux densities is plotted. 
}
\label{fig-se}
\end{center}
\end{figure*}

We note that the fluxes reported in Figures \ref{fig-time} and \ref{fig-flux} are those integrated over the whole region including both the supermaser and other features. 
We investigate possible contribution from other maser features at almost the same velocity with the supermaser feature in the 18 epochs of our VLBI imaging. 
Although it is difficult to make full channel maps with a wide field of view of an order of 10\arcsec, we search for positions of the maser features detected by \citet{gaume1998}. 
As a result, we detect one maser feature around the position offset of ($\sim$3.175\arcsec, $\sim$-1.850\arcsec) in right ascension and declination, respectively, with respect to the supermaser feature. 
\citet{gaume1998} detected two maser features near this position at the velocities of 6.3~km~s$^{-1}$ and 7.6~km~s$^{-1}$ with the flux densities of 21.0~Jy and 897.7~Jy, respectively. 
Figure \ref{fig-se} shows the flux density of this maser feature (we call this feature "SE feature") and the flux density ratio relative to the supermaser feature. 
The flux density of the SE feature increases with time and the maximum phase is observed on 2012/244. 
It is about 2 months after the maximum phase of the supermaser. 
Even around the maximum phase of the SE feature, a possible contribution to the total flux density is as large as ~8\% as shown in Figure \ref{fig-se} (b). 
Therefore, we conclude that the contamination from other maser features including the SE features are not significant in Figures \ref{fig-time} and \ref{fig-flux}. 

\subsection{Relationship between the linewidth and flux density}

\begin{figure*}
\begin{center}
\includegraphics[width=9cm]{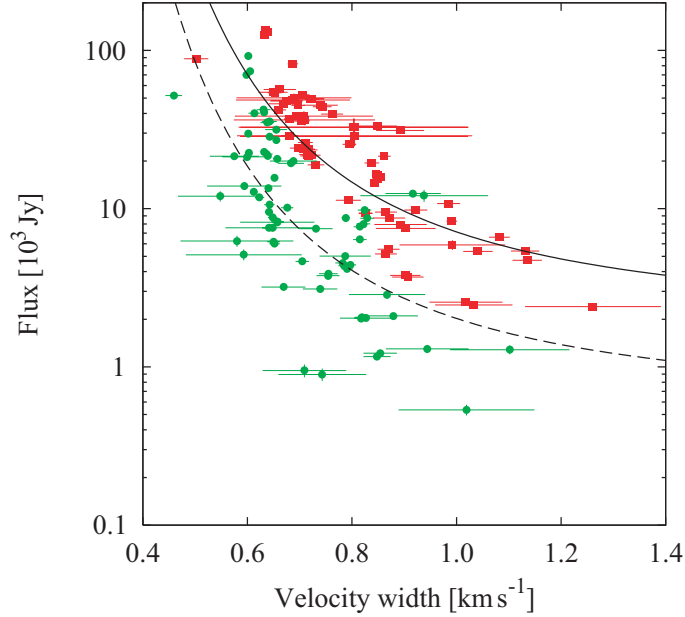}
\caption{Relationship between the flux density and linewidth of observed spectra of the supermaser. 
Red squares and green circles represent the observed values for total-power and cross-power spectra, respectively. 
Solid and dashed lines show best-fit results of the fitting; $\log F=A+B \Delta v^{-2}$, where parameters $(A, B)$ are (2.0$\pm$0.3, 1.29$\pm$0.09) and (0.58$\pm$0.16, 1.25$\pm$0.12) for the results of total- and cross-power spectra, respectively. 
}
\label{fig-dv}
\end{center}
\end{figure*}

The time variation of the linewidth of the supermaser implies an anti-correlation with the flux density as shown in Figure \ref{fig-dv}. 
According to model calculations, the line profile of the maser first tends to be 
narrower as its intensity increases as long as the maser is unsaturated \citep{goldreich1974}. 
Then the maser line starts to rebroaden when it becomes saturated \citep{goldreich1974}. 
Under the condition of an unsaturated maser, the logarithm of the flux density of the maser line, $\log F$, is predicted to be proportional to the inverse square of the linewidth, $\Delta v^{-2}$. 
In Figure \ref{fig-dv}, we also show best-fit results; 
\begin{equation}
\log F=A+B \Delta v^{-2}.
\label{eq-1}
\end{equation}

Our results may suggest that the supermaser in the current burst phase is still unsaturated. 
It is consistent with the observed narrow linewidths, 0.40-0.48~km~s$^{-1}$, of the supermaser in 1998-1999 possibly attributed to the line narrowing for an unsaturated maser \citep{shimoikura2005}. 
In contrast, no clear relationship between the flux density and linewidth is found in the burst in 1979-1985 \citep{garay1989}, suggesting a saturated maser emission. 
The difference in the degree of saturation could be attributed to the lower flux density, $\sim$10$^{5}$~Jy in the current burst phase than those in the previous bursts, $>$10$^{6}$~Jy \citep{garay1989}. 
We note that the line narrowing is predicted to occur even for a saturated maser if a stimulated emission rate is not large compared with the cross-relaxation rate \citep{goldreich1974}. 
This means that the above relation holds when the maser flux density is not enough large regardless of saturated or unsaturated cases. 
Thus, we cannot firmly estimate the degree of saturation of the supermaser by judging from the linewidth-flux density relation only. 

\subsection{Maser maps and proper motion measurements}

\begin{figure*}
\begin{center}
\includegraphics[width=18cm]{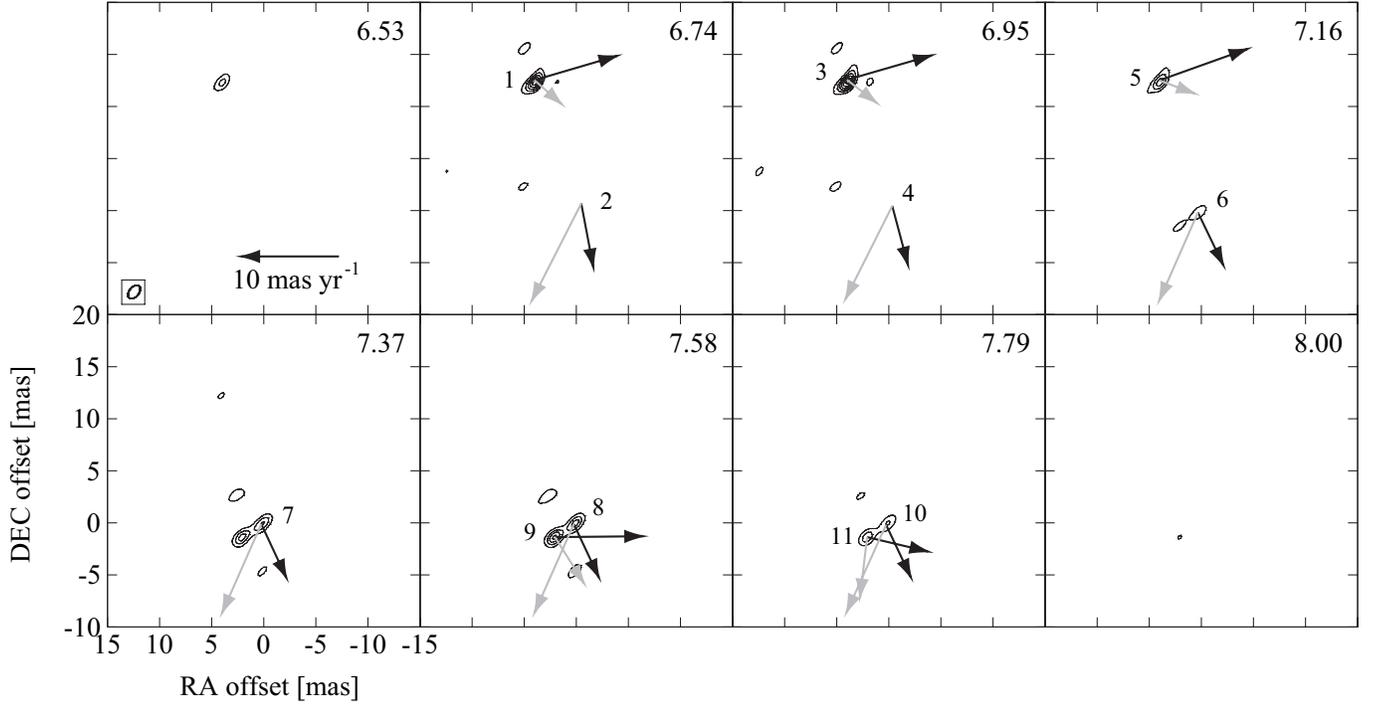}
\caption{Phase-referenced images of the bursting maser features at the epoch 2011/137. 
Absolute proper motions and proper motions with respect to Source~I are indicated by gray and black arrows, respectively. 
Contour levels are 1, 2, 3, 4, 5, and 6 $\times$2000~Jy~beam$^{-1}$. 
The delay tracking center position (0, 0) is $\alpha$=$05^{\rm{h}}35^{\rm{m}}14^{\rm{s}}.1255$ and $\delta$=$-05^{\circ}22$\arcmin36\arcsec.475 (J2000). 
The LSR velocity of each channel is indicated at the top-right corner of each panel. 
The synthesized beam pattern is indicated by an ellipse at the bottom left corner of the first panel for the 6.53~km~s$^{-1}$ channel. 
Spot IDs are indicated (see Table \ref{tab-proper}). 
Because the spots 2 and 4 are not detected on 2011/137, their positions are obtained from those at the detected epochs (2011/296 and 2011/250 for spots 2 and 4, respectively) by extrapolating their proper motions. 
}
\label{fig-map}
\end{center}
\end{figure*}

Figure \ref{fig-map} represents channel maps of the phase-referenced images of the supermaser in 2011/137. 
Three spatially distinct components are found in the initial phase of the burst from 2011/068 to 2011/137 as reported previously \citep{hirota2011}. 
It is well known that the 22~GHz H$_{2}$O line (6$_{1,6}$-5$_{2,3}$) has a hyperfine structure with 3 main components and that the $F$=7-6 and 5-4 components appear with the separation of +0.45~km~s$^{-1}$ and -0.58~km~s$^{-1}$ with respect to the central component, $F$=6-5 \citep{kukolich1969}. 
The velocity separation of -0.58~km~s$^{-1}$ is consistent with that observed for two bursting features. 
However, our results suggest that the two velocity components observed at 6.9~km~s$^{-1}$ and 7.5~km~s$^{-1}$ in the supermaser are not emitted from the same masing gas clump. 
Thus, these two velocity components are unlikely due to the hyperfine structure but they corresponds to distinct masing gas clumps with different positions and velocities. 

\begin{figure*}
\begin{center}
\includegraphics[width=18cm]{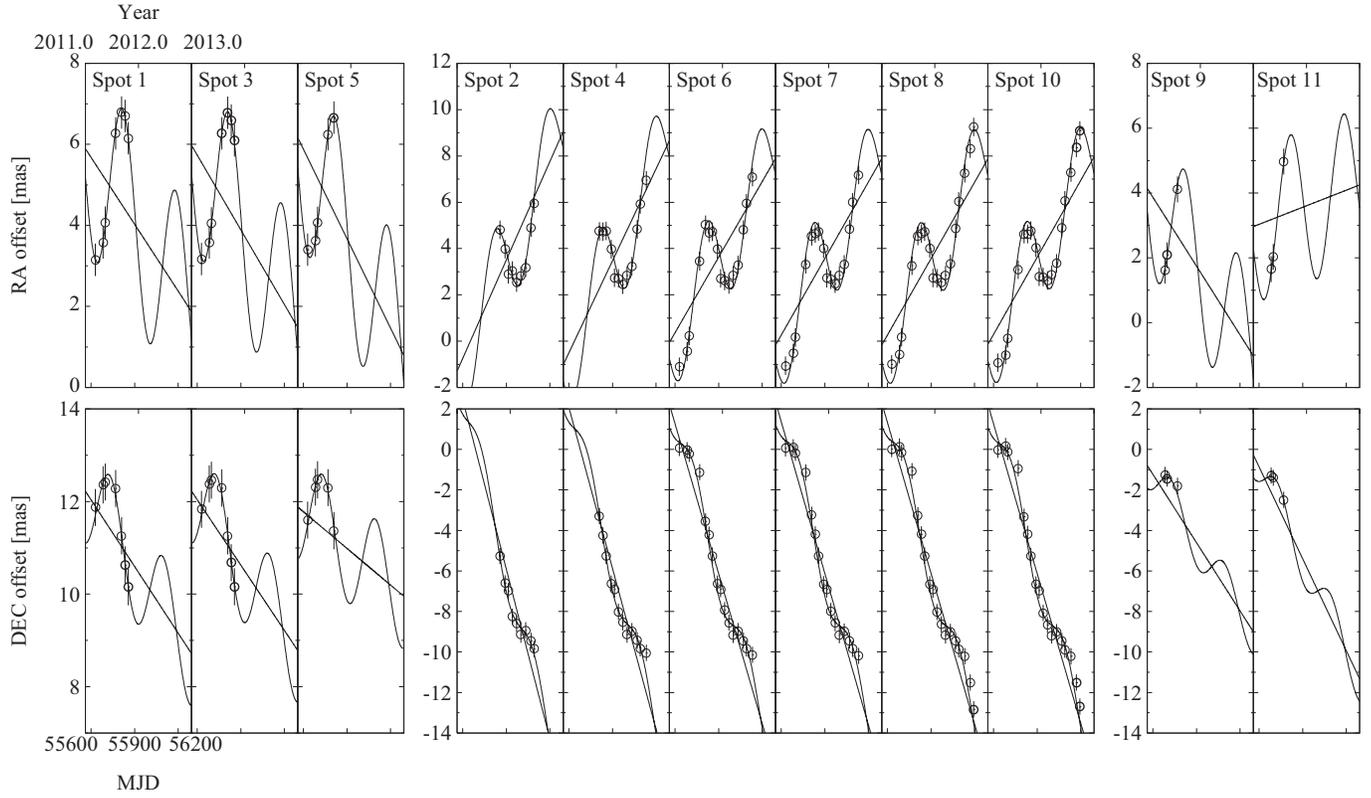}
\caption{The position offsets of the supermaser features 1 (spots 1,3,5), 2 (spots 2,4,6,7,8,10) and 3(spots 9,11) along the right ascension (top panels) and declination (bottom panels) directions as functions of time. 
A modulation of an annual parallax of 2.39~mas \citep{kim2008}, as can be seen in the sinusoidal curves, is fixed in the fitting and the proper motion of each spot, as indicated by the straight lines, is fitted (see Table \ref{tab-proper}). 
Error bars show typical positional uncertainties estimated from the post-fit residuals, 0.4~mas. 
}
\label{fig-position}
\end{center}
\end{figure*}

We conduct two-dimensional Gaussian fittings on the phase-referenced images to measure the absolute position of each maser spot. 
Hereafter, we define a "spot" as a maser emission identified in a single velocity channel and a "feature" as a group of spots that are spatially coincident in two or more consecutive velocity channels. 
By using these results, we derive the absolute proper motions of the detected spots as summarized in Table \ref{tab-proper} and Figure \ref{fig-position}. 
We find that the northern feature, feature 1 (or spots 1, 3, and 5), is detectable only until 2011/296 (Table \ref{tab-proper}). 
The lifetime of the northern feature 1 is about 8~months. 
Given the typical monitoring interval and detection criterion, in which a spot is regarded as real when it is detected for more than three contiguous epochs, we might miss a short-term variation of a maser spot on phase-referenced images within a lifetime of less than $\sim$2~months. 

According to the absolute proper motion measurements, the supermaser features are apparently moving toward south and southwest directions as shown in the gray arrows in Figure \ref{fig-map}. 
The derived proper motions are consistent but are more reliable than those previously obtained \citep{hirota2011} because we have continued the proper motion measurements in a longer observing period. 
When we subtract a modulation of an annual parallax of 2.39~mas \citep{kim2008} and the absolute proper motion of a possible powering source of the low-velocity outflow, Source~I ($\mu_{\alpha} \cos \delta$, $\mu_{\delta}$)=(6.3~mas~yr$^{-1}$, -4.2~mas~yr$^{-1}$), measured by astrometric observations with the Very Large Array (VLA) \citep{goddi2011}, the derived proper motion vectors with respect to Source~I point toward west and southwest directions as drawn in the black arrows in Figure \ref{fig-map}. 
The magnitude of the proper motions range from 2.7 to 11.4~mas~yr$^{-1}$ or from 5 to 23~km~s$^{-1}$ (Table \ref{tab-proper}). 
Similar to the previous reports \citep{shimoikura2005, hirota2011} the supermaser features are elongated along the northwest-southeast direction perpendicular to the low-velocity outflow \citep{genzel1981, hirota2007, plambeck2009, zapata2012, niederhofer2012, greenhill2013,kim2014}. 
Observed proper motions with respect to Source~I are almost perpendicular to the major axis of each feature. 

The size of the maser spot at the peak channel on the maximum epoch, 2012/164, is 2.265$\pm$0.014~mas$\times$0.950$\pm$0.006~mas (0.95~AU$\times$0.40~AU) obtained from a Gaussian fitting on the self-calibrated image. 
The peak flux density is (1.029$\pm$0.010)$\times10^{5}$~Jy, corresponding to the brightness temperature of 1.2$\times10^{14}$~K. 
The isotropic luminosity is estimated to be 2.4$\times$10$^{-4}L_{\odot}$ by assuming a linewidth of 0.6~km~s$^{-1}$. 

\subsection{Submillimeter observations with ALMA}

\begin{figure*}
\begin{center}
\includegraphics[width=9cm]{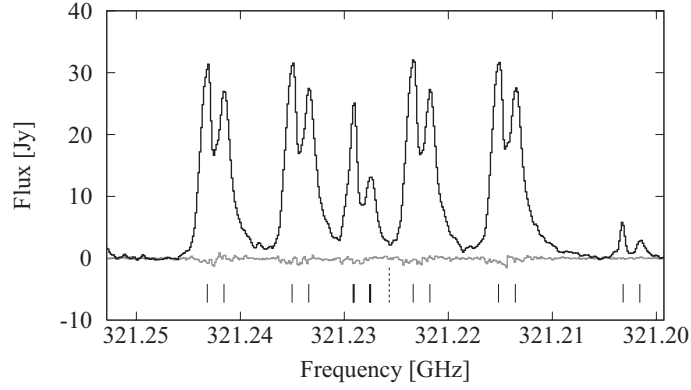}
\caption{ALMA band~7 spectra in the Compact Ridge integrated over 2\arcsec$\times$2\arcsec \ region around the supermaser. 
Black and grey lines represent spectra using full baselines and those larger than 100~k$\lambda$, respectively. 
Frequencies for each line are indicated by solid vertical lines; all show double-peaked structure at 6.9~km~s$^{-1}$ and 7.5~km~s$^{-1}$. 
The transitions and their rest frequencies are 
$^{13}$CH$_{3}$OH ($v_{t}$=0) 10$_{5,6}$-11$_{4,7}$ at 321.203~GHz, 
HCOOCH$_{3}$ ($v_{t}$=0) 28$_{3,26}$-27$_{3,25}$ E at 321.215~GHz, 
HCOOCH$_{3}$ ($v_{t}$=0) 28$_{3,26}$-27$_{3,25}$ A at 321.223~GHz,  
HCOOCH$_{3}$ ($v_{t}$=1) 30$_{0,30}$-29$_{1,29}$ A, 30$_{1,30}$-29$_{0,29}$ A, 30$_{1,30}$-29$_{1,29}$ A, and 30$_{0,30}$-29$_{0,29}$ A at 321.229~GHz, 
HCOOCH$_{3}$ ($v_{t}$=0) 28$_{2,26}$-27$_{2,25}$ E at 321.235~GHz, and 
HCOOCH$_{3}$ ($v_{t}$=0) 28$_{2,26}$-27$_{2,25}$ A at 321.243~GHz (from high to low velocity; see Splatalogue database\footnotetext{http://www.cv.nrao.edu/php/splat/}). 
The 7.5~km~s$^{-1}$ velocity component for the assumed H$_{2}$O (10$_{2,9}$-9$_{3,6}$) line at 321.226~GHz is indicated by a dashed vertical line. 
}
\label{fig-alma}
\end{center}
\end{figure*}

To study physical and chemical properties of the supermaser, we utilize ALMA cycle~0 data to make images of submillimeter continuum emissions and molecular lines at band~7. 
First, we have searched for any 321.225656~GHz H$_{2}$O line signal toward the Compact Ridge where the supermaser resides for the 321.20-321.25~GHz range, as shown in Figure \ref{fig-alma}. 
Several spectral lines are seen, but they are identified as the HCOOCH$_{3}$ lines. 
If the 321~GHz H$_{2}$O line is excited by the maser action like that at 22~GHz, it should be much more compact than those of other molecular lines. 
With this in mind, we then investigate spectra of the H$_{2}$O lines by using the visibility data with the baseline length longer than 100~k$\lambda$, corresponding to the angular size of the emission region of $<$2\arcsec. 
As a result, we can resolve out all of the lines in the Compact Ridge while a significant fraction of the 321~GHz H$_{2}$O line. 
In contrast, the broad 321~GHz H$_{2}$O line is clearly seen in Source~I with a double peaked profile \citep{hirota2014a}. 
The 321~GHz H$_{2}$O line remains in Source~I even with the baselines longer than 200~k$\lambda$ suggesting a compact structure \citep{hirota2014a}. 
Thus, we conclude that no clear evidence of the 321~GHz H$_{2}$O maser is detected at the position of the supermaser in spite of the extraordinary strong emission of the 22~GHz H$_{2}$O maser. 
We note that in the observed spectral range a vibrationally excited ($\nu_{2}$=1, 5$_{2, 3}$-6$_{1, 6}$) H$_{2}$O line at 336.227931~GHz is not detected in the Compact Ridge. 
Non-detection of the 321~GHz and 336~GHz H$_{2}$O lines implies that the physical conditions around the Compact Ridge are certainly different from those in Source~I \citep{hirota2014a}. 

\begin{figure*}
\begin{center}
\includegraphics[width=18cm]{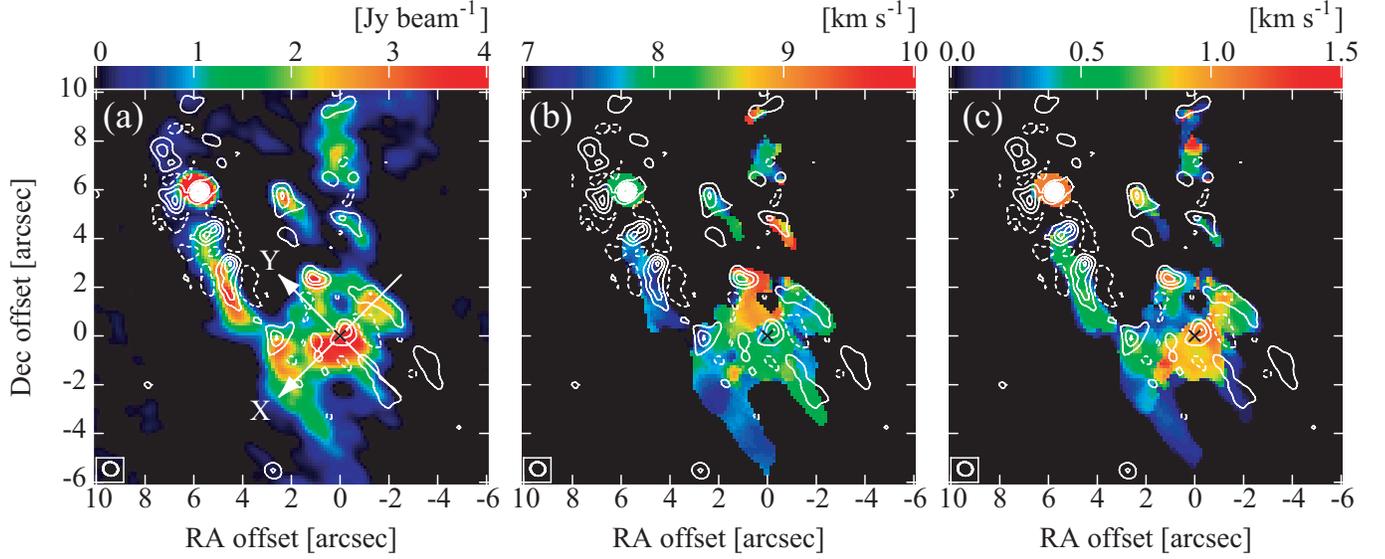}
\caption{Continuum map at ALMA band~7 (contour) superposed on moment maps of the HCOOCH$_{3}$ line at 321.223~GHz (color). 
Panels (a), (b), and (c) show the moment 0, 1, and 2 of the HCOOCH$_{3}$ line, respectively. 
The continuum map is obtained with the UV coverage larger than 100~k$\lambda$ \citep{hirota2014b}. 
The interval of contours and the lowest levels are 10$\sigma$ and $\pm$5$\sigma$, respectively, with the rms noise level, $\sigma$, of 5~mJy~beam$^{-1}$. 
A black cross at the (0,0) position in each panel shows the position of the supermaser (e.g. Figure \ref{fig-map}). 
Arrows in panel (a) indicate the X and Y axes employed in the position-velocity maps (Figure \ref{fig-pv}). 
The synthesized beam pattern is indicated by a white ellipse at the bottom left corner of each panel. 
}
\label{fig-moment}
\end{center}
\end{figure*}

We next compare the distribution of the molecular gas in the Compact Ridge which hosts the supermaser. 
For this purpose, we employ the HCOOCH$_{3}$ line at 321.22338~GHz ($v_{t}$=0, 28$_{3,26}$-27$_{3,25}$ A), which is close to the frequency of the 321~GHz H$_{2}$O line (Figure \ref{fig-alma}), as a tracer of the dense molecular gas in the Compact Ridge \citep{favre2011}. 
The 321.223~GHz HCOOCH$_{3}$ line is not blended with other HCOOCH$_{3}$ transitions unlike those at 321.229 GHz ($v_{t}$=1) lines, and hence, it is more suitable to analyze the velocity structure of the molecular gas. 
Figure \ref{fig-moment}(a) shows the moment~0 (integrated intensity) map of the 321.223~GHz HCOOCH$_{3}$ line superposed on the ALMA band~7 continuum emission. 
The supermaser is located at 8\arcsec \ southwest of Source~I and is coincident with the Compact Ridge. 
The position of the supermaser is close to the HCOOCH$_{3}$ peak but is rather located at the edge of the HCOOCH$_{3}$ condensation. 
As can be seen in the moment~1 (peak velocity) map in Figure \ref{fig-moment} (b), the velocity of the supermaser $\sim$6.9-7.5~km~s$^{-1}$ is slightly blue-shifted compared with that of the HCOOCH$_{3}$ line. 
Furthermore, the moment~2 (velocity width) map in Figure \ref{fig-moment} (c) implies that the linewidth of the HCOOCH$_{3}$ line tends to be broader around the supermaser. 
We stress that the velocity structure seems to change abruptly around the supermaser. 

\begin{figure*}
\begin{center}
\includegraphics[width=18cm]{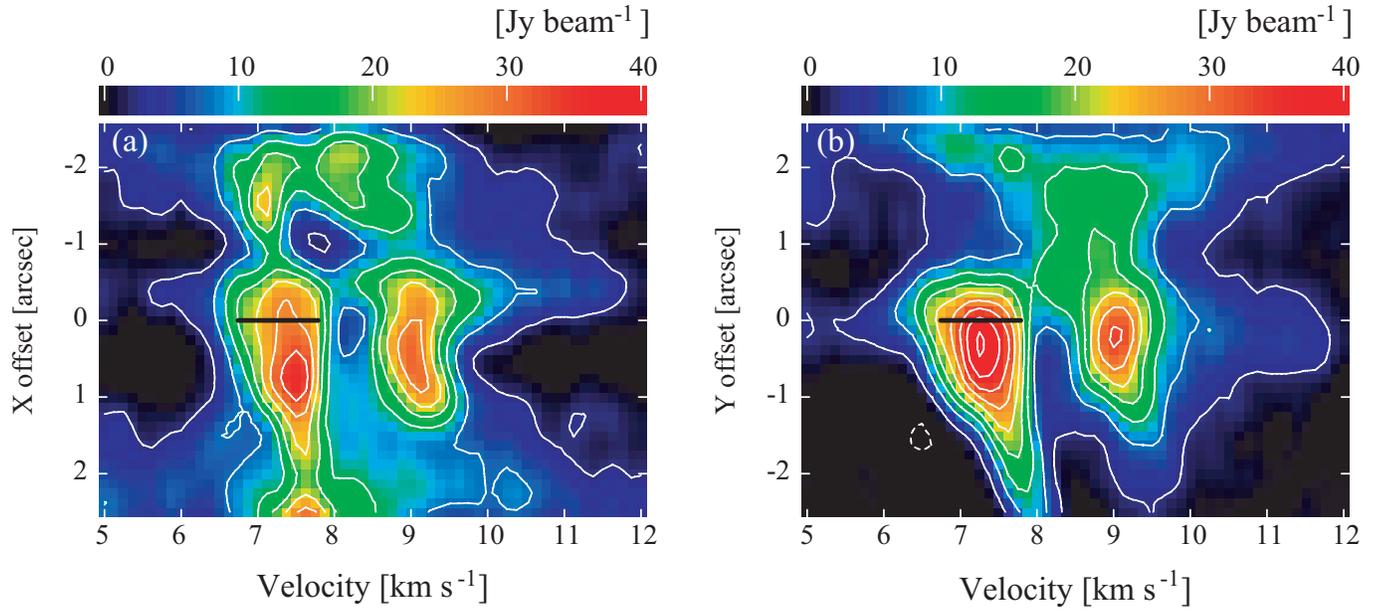}
\caption{Position-velocity maps of the HCOOCH$_{3}$ line at 321.223~GHz along the (a) X and (b) Y axes. 
The (0,0) position corresponds to that of the supermaser. 
Direction of X and Y axes are indicated in Figure \ref{fig-moment}(a). 
The interval of contours and the lowest levels are 10$\sigma$ and $\pm$5$\sigma$, respectively, with the rms noise level, $\sigma$, of 0.5~Jy~beam$^{-1}$. 
Positions and velocity ranges of the bursting H$_{2}$O maser feature identified in Figure \ref{fig-map}, 6.74-7.79~km~s$^{-1}$, is indicated by a black horizontal bar in each panel. }
\label{fig-pv}
\end{center}
\end{figure*}

These apparent velocity structures are caused by two different velocity components of the molecular gas traced by the HCOOCH$_{3}$ line.  
In Figure \ref{fig-pv}, we can see detailed velocity structures in the Compact Ridge by the position-velocity diagram along the X and Y axes as indicated in Figure \ref{fig-moment} (a).  
It is clear that there exist two different velocity components at 7.5~km~s$^{-1}$ and 9.0~km~s$^{-1}$. 
These two components were identified in the previous observations \citep{favre2011}, but we can resolve both the velocity and spatial structures by higher spectral and spatial resolutions with ALMA. 
The supermaser is thought to be associated with one of them at 7.5~km~s$^{-1}$. 
Interestingly, the velocity of the weak emission from the HCOOCH$_{3}$ line in Figure \ref{fig-pv} ranges from $\sim$6 to 12~km~s$^{-1}$, although the linewidths of the HCOOCH$_{3}$ lines are only about 1~km~s$^{-1}$ as shown in Figure \ref{fig-alma}. 
The velocity range of the weak HCOOCH$_{3}$ emission is significantly broader at X=0\arcsec \ and Y=0\arcsec, corresponding to the position of the supermaser (see Figure \ref{fig-pv}). 
This is probably an evidence of a shock front at the position close to the supermaser. 
Indeed, possible effects of shocks in the Compact Ridge have been suggested by the enhancement of organic molecules such as HCOOH \citep{liu2002}, HCOOCH$_{3}$ \citep{favre2011}, and CH$_{3}$OCH$_{3}$ \citep{brouillet2013}. 

Figure \ref{fig-cont} shows a close-up view of distributions of the ALMA band~7 continuum emission superposed on the integrated intensity map of the 321.223~GHz HCOOCH$_{3}$ line (Figure \ref{fig-moment}(a)). 
Positions of the 22~GHz H$_{2}$O masers \citep{gaume1998} including that of the supermaser are also indicated. 
We detect the compact continuum emission both at ALMA bands~6 and 7 \citep{hirota2014b}. 
The accurate position of the supermaser is significantly shifted from the continuum peak by 0.3\arcsec \ southeast. 
We also indicate positions of other possible counterparts observed previously. 
Their relationship will be discussed in the next section. 

\section{Discussion}

In this paper, we report results of continued monitoring observations of the supermaser with VERA for more than 5~years. 
We also acquire follow-up observational data with ALMA of submillimeter continuum emissions and H$_{2}$O lines at 321~GHz to reveal characteristics of the supermaser and their environments. 
In this section, we will address following issues related to the supermaser: 
What is the powering source of the supermaser and what makes the 6.9 and 7.5~km~s$^{-1}$ features in the Compact Ridge only such a special burst event?
Why there is no bursting H$_{2}$O maser feature at 321~GHz despite the extraordinary high flux density in the 22~GHz H$_{2}$O maser? 
How the drastic burst of the H$_{2}$O maser can be explained? 
Is there a periodicity of the burst event? 

\subsection{Possible powering source of the supermaser}

According to the absolute position of the supermaser derived from our VERA observations, we propose that the supermaser would be excited in the shocked molecular gas interacting between the low-velocity outflow driven by Source~I and ambient dense gas in the Compact Ridge \citep{genzel1981, hirota2011}. 
Proper motions and elongations of the supermaser features (Figure \ref{fig-map}) and the broader HCOOCH$_{3}$ line at the position of the supermaser (Figure \ref{fig-pv}) support this scenario. 
It has also been suggested by the fact that the H$_{2}$O maser tends to be distributed in the interface of the shocked molecular gas in the Compact Ridge traced by HCOOH \citep{liu2002} and HCOOCH$_{3}$ \citep{favre2011} lines. 
Furthermore, \citet{favre2011} point out a possible relation between the Compact Ridge and an explosive outflow traced by the CO lines \citep{beuther2008, zapata2009}. 
Indeed, the supermaser is found to be located at the edge of the low-velocity bipolar molecular outflow traced by the thermal SiO lines \citep{plambeck2009, zapata2012, niederhofer2012} as well as the explosive outflow \citep{beuther2008, zapata2009}. 
The latter one is thought to be the result of a recent close encounter of Source~I and BN object \citep{bally2011, goddi2011}, although a relationship between these two outflows is still unclear.  

Note that an infrared source IRc~5 is located at 1\arcsec \ southeast of the supermaser. 
The position of IRc~5 corresponds to a nearby 3~mm continuum peak, C29 \citep{friedel2011}, and an HCOOCH$_{3}$ peak, MF14 \citep{favre2011} and hence, it is more closely related to these sources. 
Since IRc~5 is thought to be a shocked molecular gas heated externally \citep{okumura2011}, it would not be directly related to the excitation of the supermaser. 
Alternatively, both of them are thought to be heated by the common outflow. 

As discussed previously \citep{hirota2011}, however, it has been still unknown why only the 6.9~km~s$^{-1}$ and 7.5~km~s$^{-1}$ features show such an extraordinary burst phenomenon. 
One of the possible explanations is a special physical or geometrical condition of the maser excitation in this region. 
An anomalous amplification of the H$_{2}$O maser burst might need a long path length. 
Such a condition could be achieved only in the systemic velocity component. 
Because it can supply plenty of ambient molecular gas within a narrow velocity range, maser emission would be efficiently amplified only close to the systemic velocity \citep{hirota2011}. 
Millimeter and submillimeter continuum emissions are detected by \citet{tang2010}, SMM1 in \citet{zapata2011} and Cb1 in \citet{favre2011} with spatial resolution of $\sim$0.8-3\arcsec. 
The mass of the submillimeter core in the Compact Ridge is estimated to be 4$M_{\odot}$ \citep{tang2010}, and hence, they could provide the dense gas required to form new stars. 
A similar conclusion is obtained from the ALMA Science Verification continuum data that have been analyzed by \citet{brouillet2014}. 

\begin{figure*}
\begin{center}
\includegraphics[width=9cm]{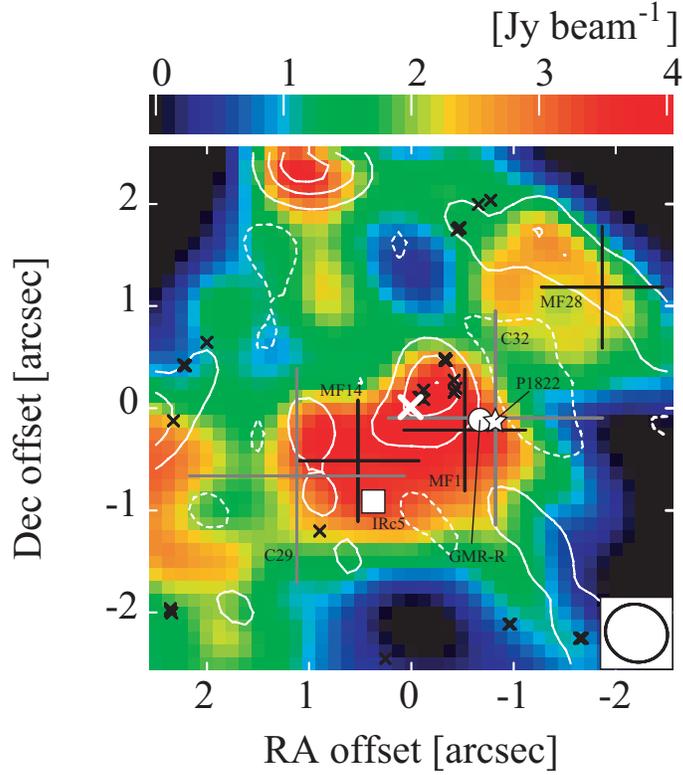}
\caption{Zoom-up of the moment 0 map of the HCOOCH$_{3}$ line at 321.223~GHz (color) and ALMA band~7 continuum map (contour) around the supermaser (see Figure \ref{fig-moment}(a)). 
The (0,0) position corresponds to that of the supermaser. 
White cross, square, circle, and star represent the positions of the supermaser, infrared source IRc5, radio source GMR-R \citep{felli1993}, and optical source Parenago~1822, respectively. 
Positions of the HCOOCH$_{3}$ peaks, MF1, 14, and 28 \citep{favre2011}, and 3~mm continuum peaks, C29 and 32 \citep{friedel2011}, are indicated by black and grey plus signs, respectively, of which sizes correspond to those of the synthesized beams. 
Black crosses represent positions of the 22~GHz H$_{2}$O masers detected by VLA \citep{gaume1998}. 
The synthesized beam pattern of the ALMA image is indicated at the bottom right corner. 
}
\label{fig-cont}
\end{center}
\end{figure*}

To account for a necessary condition for the supermaser, it is also suggested that the maser burst could occur as a result of an interaction with outflows and a pre-existing YSO in the Compact Ridge \citep{garay1989}. 
As discussed in detail by \citet{favre2011}, there are possible candidate sources physically associated with the supermaser; a radio continuum source labeled R detected with VLA \citep{felli1993}, 3~mm continuum emission sources detected with the Combined Array for Research in Millimeter-wave Astronomy (CARMA) \citep{friedel2011}, molecular peaks traced by HCOOCH$_{3}$ lines with the IRAM Plateau de Bure interferometer (PdBI) \citep{favre2011}, and optical source Parenago~1822, which is also identified from millimeter, near infrared, optical and X-ray (see SIMBAD), in the Compact Ridge (Figure \ref{fig-cont}). 

\citet{eisner2008} detect a compact 1.3~mm continuum source named HC~438 in the Compact Ridge with a flux density of 67.8$\pm$14.2~mJy by using CARMA. 
We also detect a compact continuum source at ALMA band~7 (see contours in the middle of Figure \ref{fig-cont}) which is also detected at Band~6 \citep{hirota2014b}. 
These observations achieve much higher spatial resolution of $\sim$0.5\arcsec \ than those reported above \citep{tang2010, zapata2011, favre2011}. 
The flux density at ALMA bands~6 and 7 are 57$\pm$5~mJy and 129$\pm$14~mJy, respectively \citep{hirota2014b}. 
The ALMA band~6 result is consistent with that of the CARMA 230~GHz observation \citep{eisner2008}. 
The supermaser is located within the submillimeter core detected by ALMA but is shifted by 0.3\arcsec \ from the peak position. 
Taking into account positional uncertainties, they are identical to HC~438 \citep{eisner2008}, the 3~mm continuum source C32 \citep{friedel2011} and HCOOCH$_{3}$ peak MF1 \citep{favre2011}. 
The circumstellar mass of HC~438 is derived to be 0.085~$M_{\odot}$ with the assumed dust temperature of 20~K \citep{eisner2008}. 
If we assume a higher temperature of 100~K derived from the HCOOCH$_{3}$ data \citep{favre2011}, the smaller mass of 0.01$M_{\odot}$ is obtained. 
These mass estimates would be a lower limit of the dust continuum source possibly associated with the supermaser because most of the flux is filtered out by ALMA and CARMA with extended configurations. 

We speculate that an embedded YSO associated with jet/outflow/disk \citep{matveenko1988,matveyenko2004,shimoikura2005,eisner2008} would play an important role to realize special conditions required for the supermaser. 
We notice that the H$_{2}$O maser features near the supermaser detected with VLA \citep{gaume1998} are distributed around the submillimeter continuum peak as seen in Figure \ref{fig-cont}. 
It may imply an expanding shell ejected by an embedded YSO, which are sometimes seen in massive YSOs \citep{torrelles2003}. 
As discussed in section \ref{sec-vera}, we tried to search for other maser features in Orion~KL region previously detected with VLA \citep{gaume1998}. 
We made channel maps of the H$_{2}$O maser features with VERA within an 1\arcsec$\times$1\arcsec \ field of view around the supermaser in the Compact Ridge. 
However, we found no maser emission because of the limited dynamic range seriously affected by the extremely strong supermaser. 
Further observations of continuum emission and H$_{2}$O masers with higher dynamic range will unveil the exact location of the embedded YSO in the Compact Ridge. 

\subsection{Implication for the pumping mechanism}

It is proposed that the pumping mechanism of the H$_{2}$O maser is dominated by collisional excitation \citep{elitzur1989, neufeld1990, neufeld1991}. 
The burst of the H$_{2}$O maser in Orion~KL is proposed as the result of collisional pumping in  the magnetohydrodynamic shock wave caused by the outflow \citep{garay1989}. 
On the other hand, alternative mechanisms are also considered to account for the extraordinary H$_{2}$O maser burst, such as overlapping of two maser gas clumps \citep{deguchi1989,shimoikura2005} and accidental beaming of radiation toward the Earth rather than a change in excitation \citep{genzel1981}. 
To constrain the physical properties of the supermaser, we have carried out observations of the submillimeter H$_{2}$O lines which could be masing \citep{neufeld1990, neufeld1991} by using ALMA. 
Among a number of H$_{2}$O lines, we employ the 10$_{2,9}$-9$_{3,6}$ transition at 321~GHz which is proposed to be a good indicator of some key parameters including temperature, density, and H$_{2}$O abundance in the masing gas clump \citep{neufeld1990}. 

However, there is no significant emission of the 321~GHz H$_{2}$O line at the position around the supermaser. 
The upper limit of the flux density of the 321~GHz H$_{2}$O line is $\sim$0.1~Jy. 
If we assume physical properties of the masing gas are common for both the 22~GHz and 321~GHz lines, then a photon luminosity ratio, $L_{p}$(22~GHz)/$L_{p}$(321~GHz), is identical to the observed flux density ratio, $F$(22~GHz)/$F$(321~GHz). 
Then, an upper limit of $L_{p}$(22~GHz)/$L_{p}$(321~GHz)$>$10$^{6}$ is obtained from the flux density of the 22~GHz maser of $>$10$^{5}$~Jy. 
When the beaming angle of the 22~GHz maser is also the same as that of 321~GHz, an emissivity ratio $R\equiv Q$(22~GHz)/$Q$(321~GHz) \citep{neufeld1990} is also equal to the photon luminosity ratio and flux density ratio, $R>10^{6}$. 

As demonstrated in the model calculations \citep{neufeld1990}, the lower limit of $R>10^{6}$ can be achieved when the temperature is $\sim$200-400~K or the parameter $\log \xi$ is $\sim$0.4-1.8 depending on the temperature within the calculated range below 2000~K (see their Figure 3). 
The parameter $\xi$ is defined as $\xi \equiv x_{-4}$(H$_{2}$O)~$n_{9}^{2}$/$(dv_{z}/dz)_{-8}$, in which $x_{-4}$(H$_{2}$O), $n_{9}$, and $(dv_{z}/dz)_{-8}$ are a fractional abundance of H$_{2}$O relative to hydrogen nuclei in units of 10$^{-4}$, a number density of hydrogen nuclei in units of 10$^{9}$~cm$^{-3}$, and a velocity gradient in units of 10$^{-8}$~cm~s$^{-1}$~cm$^{-1}$ \citep{neufeld1990}. 
We note that the model employed here is calculated under the assumption of a saturated maser \citep{neufeld1990} and hence, the estimated parameters may be different from an unsaturated case as adopted for the supermaser. 
Furthermore, the model calculation only considers collisional pumping of H$_{2}$O molecules ignoring infrared pumping. 
In addition, the collision rates used in these works are now superseded by newer calculations \citep{daniel2013}. 
Despite these uncertainties, we use the model of \citet{neufeld1990} to roughly interpret the supermaser burst at 22~GHz and the non-detection of the 321~GHz line. 
As a result, we find the supermaser burst at 22~GHz and the non-detection at 321~GHz could still be explained by the theoretical model \citep{neufeld1990} without extraordinary physical conditions. 
Thus, under the condition of the unsaturated 22~GHz H$_{2}$O maser, the lower temperature and larger value of $\xi$ would also be expected according to a qualitative consideration. 

To explain the non-detection of the 321~GHz maser line (high value of $R\sim \infty$), either high H$_{2}$O abundance, high H$_{2}$ density, small velocity gradient, or low temperature of 200-400~K would be required. 
The H$_{2}$O abundance in the Compact Ridge is estimated from the Herschel data to be 2.6$\times$10$^{-6}$ with the source size of 6\arcsec \ \citep{neill2013}. 
If a local abundance enhancement occurs around the supermaser, the higher H$_{2}$O abundance up to 6.5$\times$10$^{-4}$ as observed in the Hot Core region \citep{neill2013} may be possible. 
The average H$_{2}$ number density of the dust continuum source in the Compact Ridge is estimated to be (0.5-5)$\times$10$^{9}$~cm$^{-3}$ assuming a diameter (FWHM) of 0.43\arcsec \ or 90~AU \citep{hirota2014b} and a total mass of 0.01-0.1~$M_{\odot}$ \citep{eisner2008, hirota2014b}. 
If we assume a velocity width of 0.6~km~s$^{-1}$ and a path length of 1~AU as obtained from a Gaussian fitting for a bursting maser spot, the velocity gradient $(dv_{z}/dz)_{-8}$ is calculated to be 0.4. 
Although the above results contain large uncertainties by a factor of 10 or more, these parameters seem reasonable to explain estimated range of $\log \xi$ of 0.4-1.8 unless the H$_{2}$O abundance is the lowest estimate of $x_{-4}$(H$_{2}$O)=0.026 \citep{neill2013}. 

It is also conceivable that the supermaser is the result of strong beaming effects \citep{genzel1981}. 
If only the 22~GHz maser emission has a smaller beaming angle compared with that of the 321~GHz line, the emissivity ratio $R$ is proportional to the flux density ratio, $F$(22~GHz)/$F$(321~GHz), and the ratio of beaming angles, $\Omega$(22~GHz)/$\Omega$(321~GHz), which reduces the upper limit of $R$ estimated from the observed flux density ratio. 
The smaller beaming angle of the 22~GHz maser line than that of 321~GHz by a factor of $\sim$100 would enable a large $\Omega$(22~GHz)/$\Omega$(321~GHz) value of $\sim$10000. 
This may result in the lower $R$ value by a factor of 10000 which can be achieved under wider range of physical conditions as seen in Figure 3 of \citet{neufeld1990}. 

In spite of the presence or absence of a beaming effect, the condition for the H$_{2}$O maser burst only at the 22~GHz transition, $R>10^{6}$,  does not require extreme physical properties but rather those typical in hot molecular cores \citep{neill2013}. 
We suggest that the symmetric time variation in the flux density (Figure \ref{fig-flux}) could be explained by a change in the beaming angle of the maser rather than changes in the physical properties such as H$_{2}$O abundance, H$_{2}$ density, and/or temperature. 
On the other hand, a change in a velocity gradient caused by a length of the maser cloud and/or velocity structure might happen such as those in the overlapping maser model \citep{deguchi1989, shimoikura2005}. 
Nevertheless, it is still uncertain how such an accidental overlapping and/or a beaming effect could occur in two spatially distinct clouds for several times during our monitoring period of $\sim$5 years. 

\subsection{Periodicity and lifetime of the H$_{2}$O maser burst}

Comparing the basic properties such as the peak LSR velocity, the spatial structure of each feature, and the location coincident with the Compact Ridge, all of the burst events in 1979-1985, 1998-1999, and 2011-2012 could have a common origin. 
It is also suggested that the H$_{2}$O maser burst may have periodicity of 13~years \citep{tolmachev2011}. 
However, the current burst shows less activities in terms of the maximum flux density. 
Furthermore, no such burst event is reported in 1973 \citep{baudry1974}, in which the burst event is expected if 13~year-periodicity is valid (i.e. 1972-1973, 1979-1985, 1998-1999, and 2011-2012). 
In 1973 July, the observed flux density of the H$_{2}$O maser at 7.4~km~s$^{-1}$ was only about 100~Jy and was weaker than the 11.0~km~s$^{-1}$ component \citep{baudry1974}. 
This component was probably identical to the second brightest component in the current burst phase (Figure \ref{fig-sp}). 
The position of the 7.4~km~s$^{-1}$ component was measured to be R.A.=05h35m14.22s and decl.=-05$^{\circ}$22\arcmin37\arcsec.0 (J2000) and was consistent with that of the current bursting feature located in the Compact Ridge. 
In 1975-1976, the H$_{2}$O maser feature at 7.6~km~s$^{-1}$ was detected at the position of 
R.A.=05h35m14.07s and decl.=-05$^{\circ}$d22\arcmin38\arcsec.0 (J2000). 
The position of this feature was again in good agreement with that of the current H$_{2}$O maser burst, whereas it was rather weak (much less than 1000~Jy) compared with another strong feature at 10.8~km~s$^{-1}$. 

Although the H$_{2}$O maser feature at $\sim$7.5~km~s$^{-1}$ could be excited by a common origin in the Compact Ridge, it is not very likely that the maser burst occurs regularly at a period of 13~years. 
More archive data of past observations will be helpful to investigate a long-term variability of the H$_{2}$O maser activities since early 1970s. 
The flux increase started in 2013 is now on-going and hence, further monitoring will also provide some hints to understand a possible relation with previous bursts. 
Along with the 22~GHz H$_{2}$O maser lines, it could be useful to continue monitoring observations of the 321~GHz H$_{2}$O lines in Orion~KL with ALMA in order to see whether the Compact Ridge and Source~I, which is a powering source of the outflow interacting with the Compact Ridge, behave in the same way. 

Figures \ref{fig-sp}-\ref{fig-flux} imply that the current burst in 2011 and 2012 sometimes shows variations in the flux density and velocity structure with the timescales of 2-7~months. 
It is comparable to those of typical H$_{2}$O masers in Orion~KL \citep{hirota2007}. 
Although the spectral component of the supermaser seems to survive for more than 5-10~years \citep{genzel1981, garay1989}, a spatial structure of each bursting feature is found to be changing within a timescale of a few months to $<$2~years at the longest lifetime (Table \ref{tab-proper}). 
Thus, it is unlikely that the bulk of the bursting maser gas is identical for longer than $\sim$2~years. 

\section{Summary}

We have carried out astrometric monitoring observations of the bursting 22~GHz H$_{2}$O maser feature, called supermaser, in Orion~KL using VERA since 2011 March. 
Combined with spectral line data obtained as a fringe-finder, we reveal a time variation of the supermaser for more than 5~years. 
In addition, we utilize ALMA cycle~0 data at band~7 (320-330~GHz) for both spectral lines and continuum emissions to investigate physical and chemical properties of the supermaser. 
The main findings are summarized as follows: 

\begin{enumerate}
\item Within the monitoring period in 2011 and 2012, we detect three active phases with rising and falling timescales ranging from 2 to 7~months. 
We also find a smaller burst in 2009 with a slightly longer rising and falling timescales of $\sim$10~months. 
The flux variations of the supermaser seem to be symmetric in time for all of the active phases. 
\item The maximum phase of the current burst is observed on 2012/154 with the total flux density of 135000~Jy, which is one order of magnitude lower than those in previous bursts. 
Although the burst events may occur with a possible periodicity of 13~years, (i.e. 1979-1985, 1998-1999, and 2011-2012), a lower flux density by a factor of 10 in the current burst, as well as no significant burst event in 1973 \citep{baudry1974} would make the claimed periodicity of 13~years still unconvincing. 
\item The outburst of the maser emission can be seen in two velocity components at $\sim$6.9~km~s$^{-1}$ and  $\sim$7.5~km~s$^{-1}$ as reported in \citet{garay1989} and \citet{hirota2011}. 
The 6.9~km~s$^{-1}$ feature is alive only in 2011 with the lifetime of $\sim$8~months. 
Note that we detect two spatially distinct features at 6.9~km~s$^{-1}$ and 7.5~km~s$^{-1}$, and hence, these two velocity components are unlikely due to the hyperfine structure of the H$_{2}$O molecule. 
\item The flux density and linewidth show an anti-correlation as expected for an unsaturated maser emission \citep{goldreich1974}, in which the logarithm of the flux density of the maser line, $\log F$, is proportional to the inverse square of the linewidth, $\Delta v^{-2}$. 
However, we cannot rule out a possibility of a saturated maser case because such a line narrowing is also predicted to occur even for a saturated maser if a stimulated emission rate is not large compared with that of the cross-relaxation \citep{goldreich1974}. 
\item As reported in previous papers \citep{shimoikura2005, hirota2011} the bursting maser features are elongated along the northwest-southeast direction perpendicular to the low-velocity outflow driven by Source~I. 
Absolute proper motions of the bursting maser features are measured based on a longer monitoring period than that in the previous paper \citep{hirota2011}. 
Their proper motions with respect to a powering source of the low-velocity outflow, Source~I, are almost perpendicular to the elongation of the maser features, suggesting that they are related to the low-velocity outflow. 
\item The size of the maser spot at the peak channel on the maximum epoch, 2012/164, is 2.265~mas$\times$0.950~mas, and the total flux density is 1.029$\times$10$^{5}$~Jy. 
These values correspond to the brightness temperature and the isotropic luminosity of 1.2$\times10^{14}$~K and 2.4$\times$10$^{-4}L_{\odot}$, respectively. 
\item The supermaser is located at the Compact Ridge region in Orion~KL. 
Continuum peaks detected by ALMA bands~6 and 7 are located 0.3\arcsec \ northwest of the supermaser. 
As suggested by \citet{favre2011}, there are a radio continuum source labeled R \citep{felli1993}, 3~mm continuum source C32 \citep{friedel2011}, HCOOCH$_{3}$ core MF1 \citep{favre2011}, and optical source Parenago~1822, which are probably identical or closely related to the compact continuum sources detected by ALMA and the supermaser. 
\item The position of the supermaser is coincident with the dense gas traced by the HCOOCH$_{3}$ line. 
The velocity of the supermaser agrees well with one of the two velocity components of the HCOOCH$_{3}$ line detected by ALMA and PdBI \citep{favre2011}. 
A slightly broader velocity range of the weak HCOOCH$_{3}$ emission would imply a shock front at the position of the supermaser. 
\item The 321~GHz H$_{2}$O line is not detected around the 22~GHz supermaser features, although the strong maser emission is found to be associated with Source~I \citep{hirota2014a}. 
Given the high emissivity ratio, the maser burst at 22~GHz could occur under conditions of the temperature of $<$200-400~K and/or the parameter $\log \xi \sim 0.4-1.8$, suggesting a high abundance of H$_{2}$O, a high number density of H$_{2}$, and/or a small velocity gradient. 
Meanwhile, these physical properties are not extraordinary values but are very similar to typical hot molecular cloud cores. 
Thus, the supermaser burst at 22~GHz with non-detection of the 321~GHz maser emission could be explained by the theoretical model qualitatively \citep{neufeld1990}. 
Further detailed calculations would be required to interpret the observed flux ratios of the masers \citep{daniel2013}. 
\item Alternatively, the supermaser may be caused by an accidental beaming only for the 22~GHz H$_{2}$O maser \citep{genzel1981}. 
The symmetric time variation in the flux density is consistent with a result of the beaming rather than changes in physical properties of the masing gas. 
However, it still remains a matter of debate how accidental overlapping and/or beaming occur repetitively for more than $\sim$30~years since the discovery of the supermaser. 
\end{enumerate}

\bigskip
We are grateful to Alain Baudry for useful comments, which is essential to improve our scientific discussion. 
We thank the staff of the VERA stations and ALMA observatory to support our project. 
This paper makes use of the following ALMA data: ADS/JAO.ALMA\#2011.0.00199.S. 
ALMA is a partnership of ESO (representing its member states), NSF (USA) and NINS (Japan), together with NRC (Canada) and NSC and ASIAA (Taiwan), in cooperation with the Republic of Chile. 
The Joint ALMA Observatory is operated by ESO, AUI/NRAO and NAOJ. 
This research has made use of the SIMBAD database, operated at CDS, Strasbourg, France. 
T.H. is supported by the MEXT/JSPS KAKENHI Grant Numbers 21224002, 24684011, and 25108005, and the ALMA Japan Research Grant of NAOJ Chile Observatory, NAOJ-ALMA-0006. 
M.H. is supported by the MEXT/JSPS KAKENHI Grant Numbers 24540242 and 25120007.

\begin{table*}
\caption{Proper motions of the detected spots.}
\label{tab-proper}
\begin{center}
\begin{tabular}{cccccc}
\hline
Feature  & Spot     & $V_{\rm{LSR}}$ & $\mu_{\alpha} \cos \delta$ & $\mu_{\delta}$  &               \\
ID$^{a}$ & ID$^{b}$ & [km~s$^{-1}$] & [mas~yr$^{-1}$]             & [mas~yr$^{-1}$] & Epoch         \\
\hline
1       &  1   &  6.74          & -2.0(0.4) \                 &  -1.8(0.4)      & 2011/068-2011/296 \\
1       &  3   &  6.95          & -2.2(0.5) \                 &  -1.7(0.6)      & 2011/068-2011/296 \\
1       &  5   &  7.16          & -2.7(1.2) \                 &  -1.0(1.4)      & 2011/068-2011/250 \\
2       &  2   &  6.74          &  5.2(0.9)                   & -10.2(0.8) \    & 2011/296-2012/164 \\
2       &  4   &  6.95          &  4.8(0.6)                   &  -9.7(0.5)      & 2011/250-2012/207 \\
2       &  6   &  7.16          &  4.0(0.3)                   &  -9.0(0.2)      & 2011/068-2012/207 \\
2       &  7   &  7.37          &  4.0(0.2)                   &  -9.1(0.2)      & 2011/068-2012/207 \\
2       &  8   &  7.58          &  4.0(0.2)                   &  -9.1(0.2)      & 2011/068-2012/269 \\
2       & 10 \ &  7.79          &  4.0(0.2)                   &  -9.0(0.2)      & 2011/068-2012/269 \\
3       &  9   &  7.58          & -2.6(9.2) \                 &  -4.1(8.3)      & 2011/121-2011/208 \\
3       & 11 \ &  7.79          &  0.6(3.7)                   &  -5.5(3.7)      & 2011/121-2011/208 \\
\hline
\multicolumn{6}{l}{${a}$: Feature IDs are defined in the previous paper \citep{hirota2011}. }\\
\multicolumn{6}{l}{${b}$: Spot IDs are indicated in Figure \ref{fig-map}. }\\
\end{tabular}
\end{center}
\end{table*}

\end{document}